\documentstyle[preprint,aps,a4,12pt]{revtex}
\begin{document}
\draft
\hfill\vbox{\baselineskip14pt
            \hbox{\bf ETL-xx-xxx}
            \hbox{ETL Preprint 00-xxx}
%            \hbox{\today}}
%            \hbox{January 1998}}
            \hbox{March 2000}}
\baselineskip20pt
\vskip 0.2cm 
\begin{center}
{\Large\bf The choice of the symmetry group for the cuprates}
\end{center} 
\vskip 0.2cm 
\begin{center}
\large Sher~Alam$^{1}$,~M.~O.~Rahman$^{2}$,~M.~Ando$^{2}$,
~S.~B.~Mohamed$^{1}$ and T.~Yanagisawa$^{1}$
%\footnote{Permanent address: Department of Physics, University
%of Peshawar, Peshawar, NWFP, Pakistan.}
\end{center}
\begin{center}
$^{1}${\it Physical Science Division, ETL, Tsukuba, Ibaraki 305, Japan}\\
$^{2}${\it  GUAS \& Photon Factory, KEK, Tsukuba, Ibaraki 305, Japan}
\end{center}
\vskip 0.2cm 
\begin{center} 
\large Abstract
\end{center}
\begin{center}
\begin{minipage}{14cm}
\baselineskip=18pt
\noindent
%%%%%%%%%%%%%%%%%%%%%%%%%%%%%%%%%%%%%%%%%%%%%%%%%%%%%%%%%%%%%
% This is the abstract
%\begin{abstract}
 Following our recent conjecture to model the phenomenona of
 antiferromagnetism and superconductivity by quantum symmetry
 groups, we discuss in the present note the choice of 
 the classical symmetry group underlying the quantum group.
 Keeping in mind the degrees of freedom arising 
 from spin, charge, and lattice we choose the classical
 group as $SO(7)$. This choice is also motivated to accomodate
 the several competing phases which are or may be present in these
 and related materials, such as stripe phase [mesoscopically
 ordered phase], Luttinger liquids, nearly antiferromagnetic 
 Fermi liquids, charge-ordered Fermi liquids, glassy phase,
 stringy phase and perhaps more. The existence and the behavior 
 of pseudo-gap and lattice distortion are also an important consideration.
 We have lumped the charge, spin and lattice-distortion
 ordering and other orderings into the psuedogap.  
%%%%%%%%%%%%%%%%%%%%%%%%%%%%%%%%%%%%%%%%%%%%%%%%%%%%%%%%%%
% Possible motivations and rationale for these choices are 
% outlined. One of the prime motivations underlying our proposal
% is the experimental observation of {\em stripe} structure [phase]
% in high T$_{\rm c}$ superconductivity [HTSC] materials. A number 
% of experimental techniques have recently observed that the 
% ${\rm CuO_{2}}$ are rather inhomogeneous, providing evidence for 
% phase separation into a two component system. i.e. carrier-rich 
% and carrier-poor regions. 
% In particular, extended x-ray absorption fine structure [EXAFS] 
% demonstrated that these domains forms stripes of undistorted and 
% distorted local structures alternating with mesoscopic length scale
% comparable with coherence length in HTSC.
%\end{abstract}
\end{minipage}
\end{center}
\vfill
\baselineskip=20pt
\normalsize
\newpage
\setcounter{page}{2}
%------------------------------------------------------------------------ 
% Section: None
%\section{Introduction}+{others}, all in one!
%\twocolumn
% 15/6/2000	
	In a previous work one of us \cite{alam98} have advanced 
the conjecture that one should attempt to model the phenomena of
antiferromagnetism and superconductivity by using quantum
symmetry group. Following this conjecture to model the phenomenona of
antiferromagnetism and superconductivity by quantum symmetry
groups, three toy models were proposed \cite{alam99-1}, namely,
one based on ${\rm SO_{q}(3)}$ the other two constructed with
the ${\rm SO_{q}(4)}$ and ${\rm SO_{q}(5)}$ quantum groups. 
Possible motivations and rationale for these choices are 
were outlined. In \cite{alam99-2} a model to describe quantum 
liquids in transition from 1d to 2d dimensional crossover using 
quantum groups was outlined.    

	In this short note we focus on the group choice since as
mentioned before \cite{alam98} that we feel that the quantum groups
arising from the classical orthogonal groups, i.e. SO(N)
are a good and worthwhile starting point, since they
naturally incorporate the symmetry group of the insulating 
antiferromagnetic state and are naturally rich enough
to accomodate quantum liquid behaviour. 

	The main purpose of this note is make a definite choice
for the classical group underlying the quantum group.
To this end we choose $SO(7)$ for the purposes of this work. 
The choice of $SO(8)$ is also tempting from the point of
view of its octonians. 
%%%%%%%%%%%%%%%%%%%%%%%%%%%%%%%%%%%%%%%%%%%%%%%%%%%%%%%%%%%%%%%%
%Ins
% dated 14/4/2000
	We note that we may generally state that there are four 
major options confronting us towards the construction of a
model of the cuprates, viz: 
\begin{itemize}
\item{} The choice of the underlying Hamiltonian
\item{} Symmetry Group
\item{} Nature of symmetry:- for example classical
or quantum
\item{} Broken and unbroken symmeries.
\end{itemize}

	We choose a Hubbard-BCS Hamiltonian, and $SO_{q}(7)$ symmetry 
group. We take superconducting state to be dominated by a d-wave 
symmetry from experimental and theoretical considerations.
Contributions from a weak $s$ wave component can be readily
accomodated and limits may be placed on it from experiment.
Current experiments indicate that the superconducting state
is predominately $d$ wave.  

%% The choice SO_q(7):
	There are several reasons for choosing  
the classical group $SO(7)$ which underlies directly the
quantum group ${\rm SO_{q}(7)}$:-
\begin{itemize}
\item{}
As is well-known and already mentioned SO(3) \cite{alam99-1} is a 
symmetry group for antiferromagnet insulator at the level of effective
Hamiltonian. 
\item{}
On the other hand effective Hamiltonian of a superconductor may be 
described by a U(1) nonlinear sigma model [XY model], for example it 
was indicated in Ref.~\cite{don90} that the metal insulator transition may
be described by the XY model. 
\item{}
It was further pointed out in  Ref.~\cite{eme95} that superconducting 
transition on the underdoped side of oxides be described by a renormalized
classical model. In addition it is worth noting that SO(3) spin rotation and 
U(1) phase/charge rotation are symmetries of the microscopic t-J model.
\item{}
At the level of effective Hamiltonian one needs at least
a $SO(3)$ to be representative of charge ordering and
local lattice distortions. This issue is a central one as recognized 
in \cite{mos98} for example and not dealt with is charge inhomogeneity 
and phase separation.
\item{}
	In theories based on magnetic interactions \cite{and87}
for modelling of HTSC, it has been assumed that the CuO$_{2}$
planes in HTSC materials are microscopically homogeneous.
However, a number of experimental techniques have recently observed 
that the CuO$_{2}$ are rather inhomogeneous, providing evidence for 
phase separation into a two component system. i.e. carrier-rich and 
carrier-poor regions \cite{oyn99}. In particular, extended x-ray 
absorption fine structure [EXAFS] demonstrated that these domains 
forms stripes of undistorted and distorted local structures alternating 
with mesoscopic length scale comparable with coherence length in HTSC.
The neutron pair distribution function of Egami et al. \cite{ega94}
also provides structural evidence for two component charge carriers.
Other techniques also seem to point that below a certain temperature
T$^{*}$ \footnote{The following can be taken as a definition of T$^{*}$:
T$^{*}$ is an onset temperature of pseudogap opening in spin or
charge excitation spectra.} the CuO$_{2}$ planes may have ordered stripes
of carrier-rich and carrier-poor domains \cite{ega94}. The emergence
of experimental evidence for inhomogeneous structure has led to
renewal of interest, in theories of HTSC which are based on 
alternative mechanism, such as phonon scattering, the lattice
effect on high T$_{\rm c}$ superconductivity \cite{lat92,phs93,phs94,ega94}.
Polarized EXAFS study of optimally doped YBa$_{2}$Cu$_{3}$O$_{\rm y}$
shows in-plane lattice anomaly \cite{oyn99} below a characteristic 
temperature T$^{*'}$\footnote{T$^{*'}$ may be defined as follows: 
T$^{*'}$ is an onset temperature of local phonon anomalies
and T$^{*'} < T^{*}$.} which lies above T$_{\rm c}$, and
close to the characteristic temperature of spin gap opening
T$^{*}$. It is an interesting question if the in-plane
lattice anomaly is related to the charge stripe or spin-phonon
interaction. We note that it has been attempted in \cite{fuk92,fuk93}
to relate the spin gap observed in various experiments such as NMR, 
neutron scattering and transport properties to the short-range ordering 
of spin singlets.   
\end{itemize}

	Thus from the above we see that there are several phases,
namely antiferromagnetism, superconductivity, charge ordering,..
etc. If we assign $SO(3)$ to antiferromagnetism, $U(1)$ to
superconductivity and $SO(2)$ [or $U(1)$] to the pseudogap
we are naturally led to $SO(7)$. For one of the simplest group 
to embed $SO(3) \times SO(2) \times U(1)$ or alternatively 
$SO(3)\times U(1) \times  U(1)$ is ${\rm SO(7)}$. 
We note that this only one way of spontaneously breaking
$SO(7)$ there are several others.
The components of the seven-dimensional vector 
($\vec{\Phi}=(\phi_1,,,,,\phi_7)$) on which $SO(7)$ 
acts can be assigned as follows: $\phi_1=\Delta_{s}+\Delta_{s}^{\dagger}$,
and $\phi_2=i(\Delta_{s}-\Delta_{s}^{\dagger})$ for superconductivity,
$\phi_3$, $\phi_4$, and $\phi_5$ for antiferromagnetism
and $\phi_6=\Delta_{p}+\Delta_{p}^{\dagger}$,
and $\phi_7=i(\Delta_{p}-\Delta_{p}^{\dagger})$ to represent
the pseudogap. We have lumped the phenomena of charge ordering,
spin ordering and lattice distortions into the psuedogap 
since we want to see if the ordering in these systems
imply the existence of the psuedogap and temperature
T$^{*'}$. There are 21 symmetry generators for $SO(7)$.
By breaking it down as above we have accounted for 
5 generators, thus we are left with 16. These 16 generators
can be interpreted on the basis of experiments on High
T$_{c}$ materials relating to spin [neutron scattering],
charge ordering, and lattice-distortion [polarized EXAFS].
  
%%%%%%%%%%%%%%%%%%%%%%%%%%%%%%%%%%%%%%%%%%%%%%%%%%%%%%%%%%%%%%
%Now keeping in mind the discussion
%in \cite{ala98} and here we propose to model 
%a unified group for antiferromagnetic, superconducting
%and other phases in HTSC materials by ${\rm SO_{q}(5)}$.
%Even in the considerations based on SO(5) the concepts of Hopf
%maps, quaternions, Yang monopole and Berry phase \cite{zha98} 
%have crept up. This leads more support to our contention
%that model for HTSC materials must be based on quantum group
%rather than the classical group in this case ${\rm SO_{q}(5)}$
%instead of SO(5). A singlet-triplet model has been suggested
%recently by Mosvkin and Ovchinnikov \cite{mos98} in light
%of experimental considerations. This model is based on
%the Hamiltonian of the two-component spin liquid \cite{mos98}.
%The Hamiltonian considered in \cite{mos98} has SO(4)
%group symmetry. The calculations and results in \cite{mos98}
%are encouraging in that they elaborate and give insight
%into static and dynamical spin properties of cuprates
%including paramagnetic susceptibility, nuclear resonance
%and inelastic magnetic neutron magnetic scattering.
%However a central issue as recognized in \cite{mos98}
%and not dealt with is charge inhomogeneity and
%phase separation. It is tempting to replace SO(4)
%by ${\rm SO_{q}(4)}$ and see if one could account
%for charge inhomogeneity and phase separation.
	
	We now turn to classical $SO(7)$ group.
The generators of $SO(7)$ can be written as\footnote{This
relation is not particular to $SO(7)$ as is true for all
SO(N). We are making this a point so that readers not
familar with group theory may not wrongly assume to the 
contrary.}
\begin{eqnarray}
[L_{ij},L_{mp}]=i\delta_{im}L_{jp}-i\delta_{ip}L_{jm}
+i\delta_{jp}L_{im}-i\delta_{jm}L_{ip}
\label{c1}
\end{eqnarray} 
We note that it is usual to denote the Lie algebra of $SO(N)$
by $so(N)$ which is the Lie algebra of all $N \times N$ real 
antisymmetric matrices. It is useful to introduce
$N\times N$ real antisymmetric matrices $L_{pm}$
defined by
\begin{eqnarray}
(L_{pm})_{jk}=\delta_{pj}\delta_{mk}-\delta_{pk}\delta_{mj}
,~~~~~~~~~~~ p,m,j,k=1,2,....N,
\label{c2}
\end{eqnarray} 
It immediately follows from Eq.~\ref{c2} that for $p \neq m$  
$L_{pm}$ has zero elements everywhere
except for an entry $+1$ in $(p,m)$ position and
$-1$ in $(m,p)$ position, $L_{pp}=0$ and that
$L_{pm}=-L_{mp}$ [i.e. antisymmetric]. Due to the antisymmetry
one can immediately see that the the generators $L$ have
which have $7 \times 7= 49 $ elements are reduced to
$(49-7)/2=21$ elements. For $SO(N)$ the number of 
symmetry generators are $N(N-1)/2$ due to antisymmetry.  
The familiar rotation group in real three dimensions has
3 symmetry generators, this the reason why we need three
Euler angles in 3 dimensions to parametrize rotations.

It can be shown from Eq.~\ref{c2} that the generators satisfy 
the following commutation relation
\begin{eqnarray}
[L_{ij},L_{mp}]=i\delta_{im}L_{jp}-i\delta_{ip}L_{jm}
+i\delta_{jp}L_{im}-i\delta_{jm}L_{ip}
\label{c1-1}
\end{eqnarray} 
which is nothing but Eq.~\ref{c1} for the case $N=7$.

	As it is well-known any $SO(N)$ represent orthogonal rotation
in N-dimensions. Thus they leave the norm of any N-dimensional
vector invariant. If we choose an order parameter $\Phi$ with
seven components viz $\vec{\Phi}=(\phi_{1},...,\phi_7)$ we know that
quantity $\phi_1^2+...+\phi_7^2$ is invariant under $SO(7)$,
we may normalize it and set it to unity. From standard group theory we 
know that or using the definition of $L_{ij}$ we can immediately write 
down its commutation with any component of $\vec{\Phi}$, namely
$\phi_{i}$
\begin{eqnarray}
[L_{pm},\phi_{i}]=i\delta_{pi}\phi_{m}-i\delta_{mi}\phi_{p}
\label{c3}
\end{eqnarray} 
which is simply a statement of how rotation generator acts
on the $\vec{\Phi}$.

	The Kinetic energy of the system is also straightforward to write, 
since in analogy of spinning top from elementary physics it has the form 
of angular-momentum squared divided by the inertia $J^2/2I$.
Thus we can write
\begin{eqnarray}
{\cal L}_{K.E} &=&\sum_{i,j}\frac{1}{2m_{ij}} L_{ij}L^{ij}\nonumber\\
               &=&\sum_{i,j}\frac{1}{2m_{ij}}(L_{ij})^2
\label{c4}
\end{eqnarray}
where $m_{ij}$ represents the moment of inertia.
 
	The potential energy term requires some discussion
and there are several choice. We first note that it is 
straightforward to see that one can interpret $\Phi$ as
representing an order parameter and write immediately
the Landau-Ginzburg expression for free energy
\begin{eqnarray}
{\cal F}= m^2 |\Phi|^2+ \lambda  |\Phi|^4
\label{c5}
\end{eqnarray}
near the mean field transition as is familair from ordinary field 
theory and statistical physics.   
In the phenomenological picture of phase transition the 
generators of $SO(7)$ rotate the order parameter without
changing its magnitude. Now symmetry will be spontaneously
broken if a fixed direction is chosen, this happens
when for example the system settles in a particular 
phase as is well-known. On the other hand we can explicitly
break the $SO(7)$ by introducing terrms which don't
respect the $SO(7)$ symmetry, for example if we write
the following term into the potential energy
\begin{eqnarray}
{\cal L}_{P.E}^{1}=-(a^2\phi_1^2+ b^2\phi_2^2).
\label{c6}
\end{eqnarray}
This term clearly only includes  unequal contributions
from $\phi_1^2$ and  $\phi_2^2$ and thus breaks 
$SO(7)$ explicitly. It even breaks $SO(2)$ symmetry
for $a^2 \neq b^2$. If we chose $a^2 = b^2$ this would
respect $SO(2)$ symmetry whilst breaking the remainder
of $SO(7)$ symmetry. We note that it is useful to know the subgroups
for the purposes of symmetry breaking, hence we summarized some information
for the groups of interest for us namely  $SO(7)$ and $SO(8)$ 
in the appendix. Another type of term to first approximation
that we must keep in the Lagarangian [see Appendix] is
the velocity-dependent terms
\begin{eqnarray}
{\cal L}_{P.E}^{v} &\sim & \phi^{i}\phi_{i}\partial_{a}\phi_{k}
\partial^{a}\phi^{k}\nonumber\\
&\sim& (v^{a}_{ik})^{2}\nonumber\\
v^{a}_{ik} &=& \phi_{i}\partial_{a}\phi_{k}-\phi_{k}\partial_{a}\phi_{i}.
\label{c7}
\end{eqnarray}

Collecting all the terms we can write the total Lagrangian
as
\begin{eqnarray}
{\cal L} &=&\sum_{i,j}\frac{1}{2 m_{ij}}(L_{ij})^2
+\frac{\omega_{ij}}{2} (v^{a}_{ij})^{2}+{\cal L}_{ssb}
+{\cal L}_{esb}
\label{c8}
\end{eqnarray}
where ssb is shorthand for spontaneous symmetry breaking 
, esb for explicit symmetry breaking and $\omega_{ij}$
are some `velocity' parameters .

Even if we start with classical groups we end up
with quantum groups if we examine their fixed
points. This can be easily seen by examining
the connection between Kac-Moody algebra and
Quantum groups which leads to the important relation
\ref{q1} [see Appendix]. This supports the conjecture
in \cite{alam98,alam99-1} that we should start with $SO(N)_q$
since if we start with their classical counterparts
and examine their fixed points we arrive at particular
$k$ values [ which are related to $q$ values]. For
example for $SO(5)$ one arrives at $SO(5)_{k=1}$
as is obvious from the discussion in the Appendix
and as also noted in \cite{bou99}.

Another important point to note
is that non-linear sigma models have connection with
strings. In turn non-linear sigma models leads us
naturally to the notion of noncommutativity via
Kac-Moody algebra [quantum groups]. Yet another strong 
feature of quantum groups is that they unify classical
Lie algebras and topology. In general sense it
is expected that quantum groups will lead to a deeper
understanding of the concept of symmetry in physics
in particular condensed matter physics. 
 
     In conclusion, we propose $SO(7)$ as the 
classical group underlying the quantum symmetry. 
Moreover $SO(7)$ is interesting in its own right for
phenomenological studies of HTSC materials. We have
included the psuedogap into the SO(7) symmetry.

\section*{Acknowledgments}
The Sher Alam's work is supported by the Japan Society for
the Promotion of Science [JST] via the STA fellowship. 

%\section{Appendix}
%\Appendix;-does not work it is appendix!!!!!!!!!!!!!!
%\section*{}
\appendix
\section*{}

	As is well-known that orthogonal rotation in N-dimension is
specified by the $SO(N)$ groups. By definition they leave the
norm squared of $N$ dimensional vector invariant. There is
a distinction between when $N$ is odd and even, i.e. 
$SO(2n)$ and $SO(2n+1)$, in fact this is so in Cartan classification.
We recall that in Cartan classification scheme groups are
classified into the categories: $A_{n}$, $B_{n}$, $C_{n}$,
$D_{n}$,  $G_{2}$, $F_{4}$, $E_{6}$, $E_{7}$, and $E_{8}$.
For example $SU(n+1)$ are in the category $A_{n}$,
$SO(2n)$ are in $D_{n}$, and $SO(2n+1)$ are in $C_{n}$.

The maximal subalgebra of classical simple Lie algebra
of $SO(7)$ reads:
\begin{eqnarray}
SO(7) &\supset& SU(4), \nonumber\\
SO(7) &\supset& SU(2)\times SU(2)\times SU(2), \nonumber\\
SO(7) &\supset& Sp(4)\times U(1), \nonumber\\
SO(7) &\supset& G(2).
\end{eqnarray}

We note that for $Sp(4)$ the maximal subalgebra reads:
\begin{eqnarray}
Sp(4) &\supset& SU(2)\times SU(2), \nonumber\\
Sp(4) &\supset& SU(2)\times U(1), \nonumber\\
Sp(4) &\supset& SU(2).
\end{eqnarray}
$Sp(4)$ is isomorphic to $SO(5)$, and
$SO(4)\sim SU(2)\times SU(2)$. $SU(2) \supset U(1)$
and $Sp(2)$, $SO(3)$, and $SU(2)$ are all isomorphic.

The maximal subalgebra of classical simple Lie algebra
of $SO(8)$ is given by
\begin{eqnarray}
SO(8) &\supset& SU(4) \times U(1), \nonumber\\
SO(8) &\supset& SU(2)\times SU(2)\times SU(2)\times SU(2), \nonumber\\
SO(8) &\supset& Sp(4)\times SU(2), \nonumber\\
SO(8) &\supset& SU(3), \nonumber\\
SO(8) &\supset& SO(7).
\end{eqnarray}

	In a simple sense one may say that non-linear
sigma model is like a Taylor expansion in field theory.
Let us explain what this means. 
We can write the action of string [1-dimensional] propagating
in a manifold with metric $G_{\mu\nu}$ as \cite{kak91}
\begin{eqnarray}
L \sim G_{\mu\nu}(X)\partial_{a}X^{\mu}
\partial_{b}X^{\nu} g^{ab}+....
\label{a1}
\end{eqnarray}
$g_{ab}$ is the two-dimensional metric generated by
the `motion' of string in the background manifold
$ G_{\mu\nu}(X)$. A crucial observation is that
$X$'s play a dual role of coordinate of the string
in the background space and scalar field in the
2-dimensional space specifield by the metric
$g_{ab}$ [i.e. the space generated by the motion
of the string]. Different choices for the background
metrics lead to different {\em conformal field theories}.
Of interest to us is the choice that the string is
propagating on a manifold specified by a Lie 
Group[for e.g., SU(N), SO(N), etc]
in other words group manifold. We thus let $g$
be an element of the Lie group. From Eq.~\ref{a1}
we can guess that a string propagating on this group
manifold has an action of form
\begin{eqnarray}
L \sim tr(\partial_{a}g^{-1}\partial^{a}g)
\label{a2}
\end{eqnarray}
where $g$ is some function of the string field $X$.
Simple differentiation gives 
\begin{eqnarray}
\partial_{a}g=\partial_{a}X_{\mu} f^{a\mu}
\label{a3}
\end{eqnarray}
for some function $f$, thus the metric $G$ can
be expressed in terms of $f$. The exact form of
action is 
\begin{eqnarray}
S &=& \frac{1}{4\lambda^2}\int tr(\partial_{a}g^{-1}\partial^{a}g)
+k \Gamma(g)\nonumber\\
\Gamma(g) &=& \frac{1}{24\pi} \int d^{3}X \epsilon^{\alpha\beta\gamma}
tr[(g^{-1}\partial^{\alpha}g)
(g^{-1}\partial_{\beta}g)(g^{-1}\partial_{\gamma}g)]
\label{a4}
\end{eqnarray}
where $\Gamma(g)$ is the Wess-Zumino term which is integrated
over 3-dimensional disk whose boundary is two-dimensional space.
For $k=0$ it reduces to ordinary sigma-model, which is not
conformally invariant [it is asymptotically free and massive].
For special values of $k=1,2,3,..$ the theory becomes effectively
massless and has an infrared-stable fixed point at the values
of parameters $\lambda$ and $k$ related via
\begin{eqnarray}
\lambda^{2} = 4\pi/k
\label{a5}
\end{eqnarray}
Thus at these special values of $k$ we have a conformally
invariant $\sigma$ model where the theory is defined on
the group manifold. This theory is called Wess-Zumino-Witten
[WZW] model. The symmetry generators $J$ satisfy a special
case of Kac-Moody algebra, viz
 \begin{eqnarray}
[J^{a}_{n},J^{b}_{m}] = f^{abc}J^{c}_{n+m}
+\frac{1}{2}k n\delta^{ab}\delta_{n+m,0}
\label{a6}
\end{eqnarray}
In Eq.~\ref{a6} we note the following the generators $J$
carry two indices, namely $a,b,c...$ which are the Lie
group indices and $n,m..$ which are arise in the 
decomposition of the generator $J=J(z)$ in terms of
its moments, viz,
\begin{eqnarray}
J(z)= \sum_{n=-\infty}^{\infty} J_{n}z^{n-1}.
\label{a7}
\end{eqnarray}
In some sense the Kac-Moody algebra smears the generators
of ordinary Lie algebra around a circle or string.

	Finally we recall \cite{kak91,alam98} that
\begin{eqnarray}
q \leftrightarrow e^{2\pi i/(k+2)}
\label{q1}
\end{eqnarray}
If we make the above correspondence it can be shown by examining
various identities of WZW model that the braiding properties of 
WZW model at level $k$ are determined by the representation theory
of quantum groups. As a trivial check if one sets $q=1$ in
\ref{q1}, which is the limit in which quantum group reduces
to the ordinary classical group, then the right-hand side of
\ref{q1} we must set $k \rightarrow \infty$, which is precisely
the limit in which Kac-Moody algebra reduces to ordinary
classical algebra. We recall that the symmetry generators
of the WZW model obey a special case of Kac-Moody algebra.

%%%%%%%%%%%%%%%%%%%%%%%%%%%%%%%%%%%%%%%%%

	Non-linear sigma models have been extensively used
in particle theory to describe interactions phenomenologically
between strongly interacting particle. For example the
Lagarangian for non-linear sigma-model for the special case of 
$SU(2)\times SU(2)$ spontaneously broken to $SU(2)$
\begin{eqnarray}
{\cal L}= -\frac{1}{2}\frac{\partial_{\mu}\vec{\pi}
\cdot\partial^{\mu}\vec{\pi}}
{(1+\vec{\pi}^2/F^2)^2}
\label{a8}
\end{eqnarray}
where the factor $1/F$ acts as the coupling term that comes
with the interaction of each additional pion.
Expanding the expression in Eq.~\ref{a8} and keeping only
the first two terms we get
\begin{eqnarray}
{\cal L}= -\frac{1}{2}\partial_{\mu}\vec{\pi}\cdot\partial^{\mu}\vec{\pi}
+(1/F^2)\vec{\pi}^{2}\partial_{\mu}\vec{\pi}\partial^{\mu}\vec{\pi}+....
\label{a9}
\end{eqnarray}
The first term is the simple kinetic energy term, the second
being the potential energy term of the form 
${\pi}_{i}{\pi}^{i}\partial_{\mu}\vec{\pi}_{j}\partial^{\mu}{\pi}^{j}$
[in words it is a velocity dependent potential term].
We must also retain in general in our phenomenological modelling
[as in this paper] of HTSC material keep such terms.

%    In conclusion, the quantum orthogonal groups 
%${\rm SO_{q}(N)}$ are proposed as potential candidate 
%for modelling the theory of HTSC materials. A strong
%feature of quantum groups is that they unify classical
%Lie algebras and topology. In more general sense it
%is expected that quantum groups will lead to a deeper
%understanding of the concept of symmetry in physics. 
 
%\end{itemize} 
%%%%%%%%%%%%%%%%%%%%%%%%%%%%%%%%%%%%%%%%%%%%%%%%%%%%%%%%%%%%
%\begin{enumerate}
%\item{}
%\end{enumerate} 
%%%%%%%%%%%%%%%%%%%%%%%%%%%%%%%%%%%%%%%%%%%%%%%%
%In a metallic system close to a metal-insulator transition
%the electronic properties play an important and a crucial role.
%%%%%%%%%%%%%%%%%%%%%%%%%%%%%%%%%%%%%%%%%%%
%\newpage

%\begin{thebibliography}{99}

%\end{thebibliography}
\end{document}